# Speed of sound of pure water to 700 MPa and an equation of state to 2300 MPa.

Olivier Bollengier[1], J. Michael Brown[1], George H. Shaw[2]

[1] Department of Earth and Space Sciences, University of Washington, Seattle, WA 98195, United States
[2] Geology Department, Union College, Schenectady, NY 12308, United States

## Abstract

Sound speeds of pure fluid water are reported between 0.1 and 700 MPa, from 353 K down to the melting curves of the ice phases. The $2\sigma$ precision and accuracy of the new sound speed measurements are close to 0.02%, with an estimated pressure accuracy of 0.04% up to 700 MPa. Using additional published measurements, a new equation of state is derived extending from 240 to 500 K and from 0.1 to 2300 MPa, covering much of the sub-critical domain of water up to the ice VI – ice VII transition. Analysis of measurements and construction of the equation of state are accomplished with a flexible computational thermodynamic framework based on local basis functions in the form of tensor B-splines. Relative to IAPWS-95 (the most comprehensive representation available), improvements in the accuracies of density, sound speed and specific heat are observed above 100 MPa, particularly near the solid-fluid phase boundaries. Accurate thermodynamic properties for aqueous solutions under an increasing range of pressures, temperatures and compositions are needed to address a variety of technical and scientific challenges: the present study provides measurements essential for improving the high-pressure and low-temperature representation of water.

# 1. Introduction

From energy engineering to resource extraction, waste treatment and the chemical industry, the use of water and aqueous solutions under an increasing range of pressures and temperatures is central to some of the most critical needs of our societies (Wagner and Pruß, 2002). Water is the enabler of life on Earth, as it governs our rich biochemistry (Ball, 2017) and takes part in geological processes shaping the habitability of our planet (Grove et al., 2012; Korenaga, 2012; Kaltenegger, 2017). Water is a major building block of moons and dwarf planets of the outer solar system, and similarly, water-rich exoplanets may be expected to abound (Mulders et al., 2015; Monteux et al., 2018). For this large population of worlds, aqueous reservoirs at pressures of hundreds to thousands of MPa appear as an increasingly common feature (Hussmann et al., 2015; Nimmo and Pappalardo, 2016). Knowledge of the properties of liquid water is thus required to address societal issues and to understand the geological behavior and habitability potential of planetary bodies throughout the universe.

Following a computational thermodynamics approach, Wagner and Pruß (2002) published a representation for the Helmholtz energy of water (a function of density and temperature) for the International Association for the Properties of Water and Steam. This IAPWS-95 formulation relies on an assembly of various global and partially localized functions to reproduce selected datasets, primarily from the melting curves to 1273 K, at pressures up to 1000 MPa. Near ambient conditions where highly accurate reference data are available, IAPWS-95 reproduces the density of pure water to 1 ppm, its sound speed to 50 ppm, and its heat capacity to 1000 ppm (0.1%). However, limited data at higher pressures prevent this level of accuracy to be sustained, and above 100 MPa, at temperatures near the melting curves of ices III, V and VI, these uncertainties are estimated to reach 0.2% for the density, 0.3% for the sound speed, and 20% for the heat capacity (IAPWS R6-95). Since the release of IAPWS-95, new studies have improved our knowledge of thermodynamic properties in sub-regions of its domain, and data on supercooled water has been used to establish an equation of state covering the metastable region below the melting curves, down to the homogenous nucleation temperature (Holten et al., 2014). However, the complexity of the IAPWS-95 formulation has hindered its modification to account for these works.

To explore the detailed thermodynamic behavior of both pure water and aqueous solutions at high pressure, we developed a new apparatus for the measurement of the speed of sound of fluids from 0.1 to 700 MPa, at temperatures between 253 and 353 K. The present publication focuses on pure water, both as a standard to assess the accuracy of our observations (prior accurate sound speed measurements exist within regions of our domain of coverage), and as an end-member composition for further studies of aqueous solutions.

For pure water, near ambient pressure, the most accurate (10 ppm) sound speeds were already included in the IAPWS-95 formulation (Del Grosso and Mader, 1972; Fujii and Masui, 1993); comparable measurements (accurate to 30 ppm) were reported more recently (Li et al., 2016). Since IAPWS-95, improved sound speed measurements under pressure have been reported. At intermediate pressures, Benedetto et al. (2005) have explored temperatures from 274 to 394 K, up to 90 MPa, with accuracies of 500 ppm. Lin and Trusler (2012) have measured sound speeds at pressures up to 400 MPa, down to the melting temperature, with estimated uncertainties of 300 ppm. Vance and Brown (2010) and Baltasar et al. (2011) have explored a larger pressure range, to 700 MPa, but with larger uncertainties (0.3% and 0.2%, respectively). Sound speeds measured in diamond anvil cells span a higher-pressure regime almost to 10,000 MPa, at temperatures to more than 700 K, but at the cost of a lower accuracy (Abramson and Brown, 2004; Decremps et al., 2006; Sanchez-Valle et al., 2013).

To analyze the sound speed measurements and establish a new equation of state for fluid water, we use the numerical framework given by Brown (2018) for the construction of thermodynamic representations. This approach, using local basis functions (LBF) in the form of tensor B-splines ($6^{th}$-order piece-wise polynomials on small intervals), removes the need for a search for problem-specific combinations of custom basis functions as done in previous studies. Analytic derivatives and integrals of LBFs allow straightforward determinations of thermodynamic properties based on any energy potential. Extension of the domain of coverage to include new or revised measurements can be accomplished without impacting the quality of the fit in the initial domain: new datasets may thus be added, and the representation updated, all using minimal computing power. Here, for its convenience in working with measurements, the Gibbs energy (a function of pressure and temperature) is used.

Using the LBF framework, the new sound speed measurements, combined with other available thermodynamic properties, are used to produce a new equation of state of pure water covering the pressures from 0.1 to 2300 MPa (up to the liquid – ice VI – ice VII triple point) for temperatures between 240 and 500 K. In Section 2, we describe the apparatus and the associated metrology, the data acquisition strategy, and the resulting measurements. This is followed in Section 3 by a description of the computational approach used to develop our equation of state. A Gibbs energy surface constructed to match available data is presented. In Section 4, we discuss the improved density and specific heat representations resulting from our new data and equation of state. We finally engage in error analysis and compare our equation of state with the IAPWS-95 standard to establish a summary assessment of the accuracy of our equation of state and experimental setup as a reference for future works.

# 2. Sound speed measurements

## 2.1. Apparatus

Pressures up to 700 MPa are generated in a Harwood steel pressure vessel using kerosene as the hydraulic fluid. The vessel is sealed by a grade 5 titanium alloy end closure supporting the elements holding the fluid sample, all machined from pure titanium for corrosion resistance (Figure 1). The acoustic chamber consists of a centimeter-long hollow cylinder attached to the end closure on one side and an acoustic reflector on the other (Figure 2). These parts are held together with two bolts tightened to a specific torque to ensure a reproducible mechanical assembly. An overlying bellows assembly (containing the sample fluid during the experiments) is also attached to the end closure. Radial holes in the hollow cylinder ensure the free circulation of the fluid during changes in pressure and temperature.

An ultrasonic transducer is attached to the end closure outside the pressure vessel. An aluminum cooling block circulating water at 282 K maintains the transducer within its operational range. For temperature regulation, the pressure vessel is surrounded by copper tubing connected to a Lauda E100 / RE107 fluid chiller working between 245 and 360 K. The vessel and copper tubing are insulated with thick external layers of aerogel blanket, fiberglass and polyethylene. Temperatures are measured close to the pressurized sample with a thermocouple in contact with the end closure at the base of its buffer rod (location "N", Figure 1).

## 2.2. Metrology

### 2.2.1. Speed of sound

Measured travel times of acoustic waves within the sample chamber of known length give the sound speeds. The ultrasonic transducer attached to the end closure sends a frequency-modulated pulse centered around 5 MHz (a domain of non-dispersive frequencies for water) through the buffer rod and to the sample chamber. The waves travelling through the end closure are partially reflected and partially transmitted at the front and back walls of the sample: the time delay between reflections from these two interfaces corresponds to the round-trip travel time within the sample cavity. Repeated reflections in the sample chamber result in a series of echoes of decreasing intensity, all multiples of the sample travel time. The reflections are collected by the transducer and recorded using a digital oscilloscope. The collected waveform is deconvolved using the source signal as the reference, and the travel time is determined from the series of sample reflections to ± 0.9 ns (Supplementary Material A, 1.1).

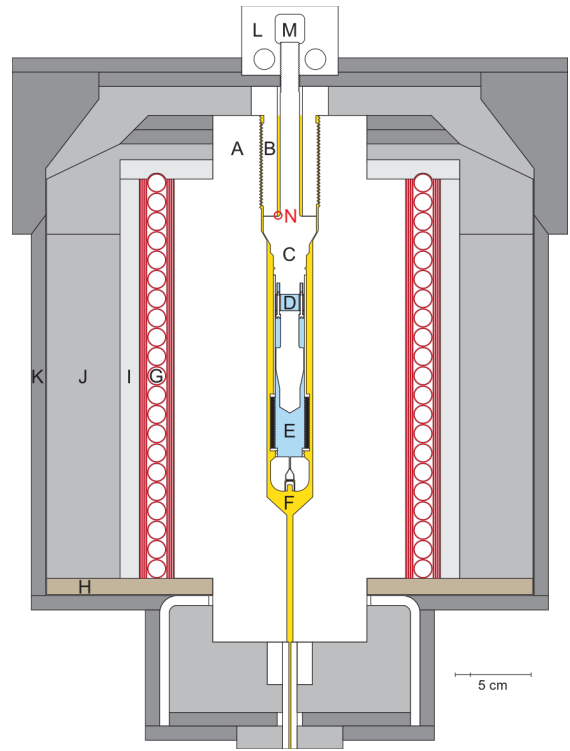

*Figure 1. Pressure vessel loaded for an experiment. A: pressure vessel. B: closing screw. C: end closure (Bridgman seal not represented). D (blue): acoustic chamber. E (blue): extra sample volume in the bellows area. F (yellow): hydraulic fluid. G: heat regulation circuit. H (brown): rock fragments for support and insulation. I (light gray): aerogel blanket. J (medium gray): mineral wool. K (dark gray): polyethylene mat. L: transducer cooling block. M: ultrasonic transducer. N (red circle): proxy location for sample temperature measurement with a thermocouple.*

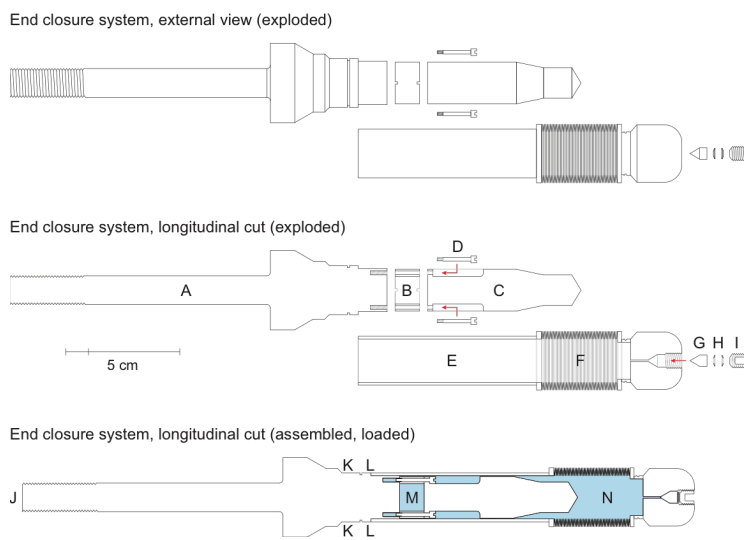

*Figure 2. End closure and sample holder. On all three views, black lines represent external, visible edges. For the cut views, gray lines represent major out-of-plane edges. A: end plug and buffer rod. B: hollow cylinder / sample chamber. C: end reflector. D: screws. E: bellows tubing section. F: bellows rings. G: cone plug. H: opposing Belleville washers. I: set screw. J: sitting surface for the ultrasonic transducer (not represented). K: sitting area for the Bridgman seal (not represented). L: sitting area for the silicone seal (not represented). M (blue): filled sample chamber. N (blue): extra sample volume in the bellows area. Holes in the sample chamber wall (B) allows the fluid sample to flow freely between the sample cavity (M) and the extra sample reservoir (N) upon changes of pressure and temperature.*

The individual path-length for each experiment was obtained by comparing our results near 0.1 MPa with the IAPWS-95 sound speeds. At this pressure, between 273 and 368 K, the reference formulation is accurate to 50 ppm. For each experiment, a sample length was searched to bring our sound speeds in agreement with the IAPWS-95 values above 273 K. Lengths for the six samples (each requiring a mechanical disassembly and re-assembly of the end closure and sample chamber) were found between 10767.5 and 10769.5 µm, with a resolution limited to 0.5 µm due to the scatter of our data. As an independently confirmation, the length of the cylinder forming the sample chamber was measured with a micrometer under ambient conditions at 10768 µm, with a resolution of 1 µm. The compressibility (Fisher and Renken, 1964) and expansivity (Hidnert, 1943) of titanium are used to adjust the reference sample length at pressures and temperatures departing from the standard conditions.

### 2.2.2. Pressure

Two pressure transducers were installed on the hydraulic circuit, each connected to a separate meter. The background noise of each transducer-meter system was measured at ± 0.015 MPa. Two additional sources of error affecting our pressure readings were apparent at room pressure: a linear dependence of the transducers output to their temperature of operation, and a long-term (hundreds of hours) drift of unknown origin. The transducers were thermally insulated and their variations in temperature measured and compensated for, while daily reference values at room pressure were acquired to compensate for the drift. After compensation of both effects, residual random errors on the pressure measurements are observed at ± 0.04 MPa (Supplementary Material A, 2.1).

One of our meter/transducer system was calibrated at Harwood Engineering against a DWT-1000-200K deadweight tester (Newhall et al., 2003). The calibration was performed along a pressure decrease cycle to follow our regular experimental protocol (Section 2.3.2), at pressures near 689, 552, 414 and 276 MPa, with a stated relative accuracy of 0.02%. Interpolation is provided by a polynomial fit starting at 0.1 MPa, with an adopted 2σ relative accuracy of 0.04% over the whole pressure range (0.28 MPa at 700 MPa). Using the speed of sound of water as a reference, the mid-range, full scale hysteresis of our experimental system has been estimated at 0.6 MPa (Supplementary Material A, 2.2). All the sound speed measurements reported in this publication have been acquired along decreasing pressure steps to remove this hysteresis as a source of uncertainty (Section 2.3.2). The reproducibility of our measured sound speeds spanning six loads over two years provides evidence that no significant drift of the pressure sensors occurred. The difference between the measured and sample pressures resulting from the spring force of the bellows was found to be negligible (Supplementary Material A, 2.3).

### 2.2.3. Temperature

Sample temperatures are measured at a proxy location by a thermocouple continuously calibrated against an RTD, as described below. We assign to our temperature measurements a $2\sigma$ accuracy of 50 mK.

Two Type T ungrounded thermocouples and a 4-wire Pt100 RTD are connected to a National Instruments module operated inside a thermally insulated box. The RTD has a baseline noise of $\pm$ 3 mK. Daily calibrations of the acquisition line, using high precision resistors, showed variations below 0.5 m$\Omega$, affecting readings to $\pm$ 1 mK. The custom calibration of the RTD was checked in a deionized water ice bath and found accurate to $\pm$ 2 mK (Supplementary Material A, 3.1). The two Type T thermocouples have a baseline noise of $\pm$ 10 mK, with longer-term fluctuating errors of $\pm$ 100 mK. The long-term fluctuations were found to be fully correlated, and are thus believed to result from the data acquisition electronics rather than being intrinsic to the thermocouples. Monitoring of the thermocouples readings in deionized water ice baths and in a temperature-regulated oil bath between 243 and 363 K revealed parallel behaviors in time with an offset of no more than 20 mK (Supplementary Material A, 3.2).

The sample temperature is measured at location "N" in Figure 1 using one of the thermocouples. The narrow space hosting the "sample" thermocouple is filled with oil to enhance thermal contact and prevent perturbation of the measurements by atmospheric circulation. The second, "control" thermocouple is set alongside the RTD in the temperature-regulated bath of the Lauda fluid chiller, at a temperature close to that of the sample. At any time, the RTD determines the absolute temperature of the control thermocouple and by extension provides the necessary correction for the long-term fluctuations of the sample thermocouple. This configuration provides the accuracy of the RTD to the thermocouple measurements at the proxy location, limited only by the precision of the thermocouples. The combined errors noted above suggest an absolute error below 50 mK in the determination of the sample temperature.

Using a mock-up end closure with a small through-hole to allow the insertion of a thermocouple inside the sample cavity, no difference in temperature between the sample cavity and the sample temperature proxy location could be resolved over our temperature range of 253 – 353 K. A finite-element calculation of steady-state temperatures based on the geometry of the pressure vessel and surrounding insulation supported this observation.

This calculation of the limit on the error in temperature is corroborated by the observation that our sound speeds agree within ± 100 ppm with the values of IAPWS-95 at 0.1 MPa between 273 and 353 K, which requires a temperature error no larger than 50 mK.

### 2.2.4. Equivalent sound speed precision

The overall precision of the sound speeds determinations is constrained by uncertainties on the time, length, pressure and temperature measurements. A summary of the four sources of random errors described in this section, with their propagated effect on the sound speeds as described below, is presented in Table 1. Gradients of the IAPWS-95 sound speed with respect to pressure and temperature were used to calculate the effect of the random errors in pressure (± 0.055 MPa) and temperature (± 0.034 K) on the sound speed determinations between 253 and 353 K, from 0.1 to 700 MPa. The uncertainties in travel time (0.9 ns) and length (0.5 μm), using a reference sample length of 10768 μm, were used to establish similar error maps. The four contributions are combined to produce a global estimate of sound speed determination uncertainties. The resulting 2σ precision is of 175 to 295 ppm. As shown later, misfits of sound speeds based on our Gibbs energy representation are consistent with these estimates. The effect of the systematic errors in pressure and temperature on the equation of state are explored in Section 4.3.

| SOURCES OF RANDOM ERRORS | MAGNITUDE | EQUIVALENT SOUND SPEED |
|---|---|---|
| TRAVEL TIME | 0.9 ns | 60 – 105 ppm |
| SAMPLE LENGTH | 0.5 μm | 45 ppm |
| PRESSURE | 0.055 MPa | 20 – 70 ppm |
|    Background noise | 0.015 MPa | |
|    Time/temperature correction residual | 0.04 MPa | |
| TEMPERATURE | 34 mK | 5 – 125 ppm |
|    RTD background noise | 3 mK | |
|    RTD daily calibration | 1 mK | |
|    Thermocouple noise | 10 mK | |
|    Thermocouple dT | 20 mK | |

*Table 1. Summary of random errors and their effect on the precision of the sound speed determinations.*

## 2.3. Protocol

### 2.3.1. Loading and sealing

Prior to each load, the end closure, hollow cylinder, acoustic reflector and screws are cleaned, rinsed with deionized water, dried, and kept dust-free until reassembly. The bellows is cleaned with a water-ethanol mixture using a soft brush, rinsed with deionized water, and dried at 500 K for an hour. At the time the bellows section is mated with the end closure, the area of contact between the end closure and the bellows is covered in silicone caulk to provide a water-proof seal. Outside the bellows, the fresh silicone is overlaid with a layer of aluminum foil glued during curing. After curing of the silicone, the aluminum layer is wrapped in a thin copper sheet (for protection against tearing) and clamped with a spring steel ring to keep the whole seal tight. Finally, the spring ring is covered with a large band of PTFE tape to further shield the remaining exposed sections of silicone from the pressurizing fluid.

Water is loaded through the conical aperture at the end of the bellows. A system of tubes and valves connects the bellows, a reservoir containing the sample to be loaded (deionized water), and a vacuum pump. The bellows is checked for mechanical integrity by its ability to hold vacuum before the fluid sample is allowed to fill the assembly. The conical filling aperture is plugged using a titanium cone of matching shape kept in place by two Belleville washers loaded by a set screw.

### 2.3.2. Data acquisition strategy

Owing to the large thermal inertia of the pressure vessel, runs were carried out along daily isotherms, with the temperature of the system changed overnight. To increase pressure stability and mitigate the hysteresis of our pressure sensors, all isotherms were explored with decreasing pressure steps. Between measurements, a delay of about 1 minute per 4 MPa of pressure change allowed adequate thermal equilibration. The pressure and temperature measurements associated with each ultrasonic acquisition were based on sensor averages over about 1 minute (± 10 points around the sound speed acquisition time, at 1000 S/h) to mitigate noise-level fluctuations. During post-acquisition review, any measurement not meeting the minimum delay interval or with a standard deviation for acquired pressures of more than 0.2 MPa was culled from the data-set. About 10% of the measurements did not meet this quality standard and were not included in the reported table.

To check for possible contamination of the sample during a run, sound speed measurements were performed at identical conditions at the beginning (before pressurization) and at the end (after returning to 0.1 MPa) of each run. Early attempts with prototype seals resulted in the samples being contaminated with the pressurizing fluid. This was obvious from a lack of sound speed reproducibility and through examination of the recovered samples. With the proper seal, runs could be carried out successfully over up to four weeks of daily pressure cycles with no change in the sound speed within measurement precision.

## 2.4. Results

Our final dataset of 901 sound speed measurements from six pure water loads, along isotherms between 253 and 353 K, at pressures from 0.1 to 700 MPa, is presented in the supplementary materials (Supplementary Material B) and in Figure 3. To assess our reproducibility, some isotherms were observed with several loads; the isotherm at 313 K was explored with each load. Measurements were acquired down to the melting curves of ices Ih, III, V and VI.

The study of Lin and Trusler (2012) is the only one providing sound speed measurements of water with a precision comparable to ours and with a documented pressure calibration covering a major part of our pressure range. There are eight isotherms for which measurements are available from both studies to establish a comparison. The pressures associated with the sound speeds reported by Lin and Trusler are based on their initial pressure calibration to 400 MPa. Their final pressure calibration, limited to 260 MPa, was used only to estimate the accuracy of their adopted pressure scale. We compared the sound speeds of Lin and Trusler using both their initial pressure calibration (published values) and final pressure calibration (Trusler, personal communication) extrapolated linearly to 400 MPa. The difference between the two pressure scales translates into a linear deviation of the sound speeds reaching about 200 ppm at 400 MPa. A similar systematic deviation was observed between our dataset and the published sound speeds of Lin and Trusler. Consequently, the Lin and Trusler sound speeds referred to in the present study are based on the revised pressures (final pressure calibration) providing a closer match to our own measurements.

Measurements along the eight isotherms common to Lin and Trusler (2012) and this study are compared with predictions of IAPWS-95 in Figure 4. The IAPWS-95 sound speeds at 0.1 MPa, with an accuracy of 50 ppm, constitute an appropriate standard to assess the quality of both datasets at low pressure. The close agreement between our dataset and that of Lin and Trusler, each based on independent pressure calibrations, supports the complex pattern of deviations from IAPWS-95 apparent Figure 4. The observed deviations from IAPWS-95 match its stated accuracy: within 0.03% above 300 K and below 200 MPa, close to 0.1% below 300 K and 100 MPa, and within 0.3% at higher pressures.

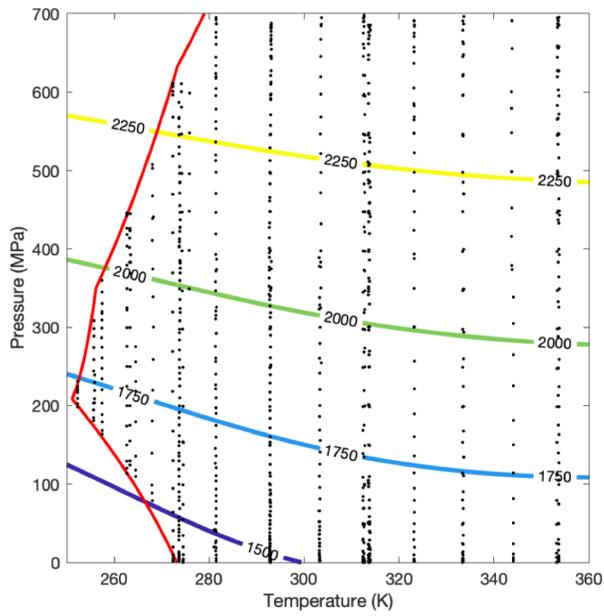

*Figure 3. New sound speed dataset. Each black dot represents an individual ultrasonic measurement. The blue, green and yellow lines are contours of a surface fit to the data (with sound speed values in m/s). The red lines are the pressure-temperature locations of the melting curves of ices Ih, III, V and VI.*

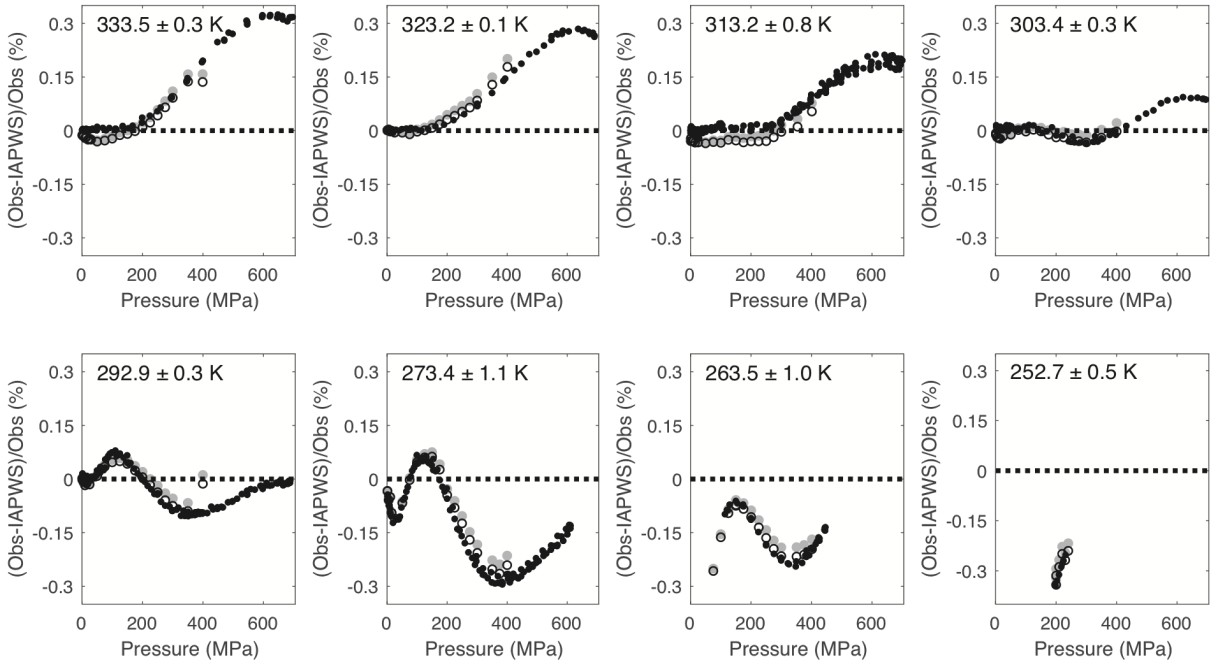

*Figure 4. Deviation of measured sound speeds from IAPWS-95. Black dots: this work. Open circles: Lin and Trusler (2012), modified pressures (as used in the present study). Gray Circles: Lin and Trusler (2012), published pressures (presented here for comparison). The eight temperatures shown are the isotherms for which data are available from both studies. The temperature variation for each isotherm is indicated by the given spread of reported temperatures. Two other sound speed datasets (Vance and Brown, 2010; Baltasar et al., 2011) covering the same pressure range to 700 MPa have uncertainties as large as the full scale of these figures and are not reported here.*

# 3. A new equation of state

## 3.1. Overview

Using methods described in Brown (2018), an analytic Gibbs energy representation is created for water based on local basis functions (LBF). As previously demonstrated, accurate sound speeds on a dense grid of pressures and temperatures allow replication of an underlying equation of state to ppm levels. Here, we extend this approach using a variable distribution of sound speed measurements. Additional constraints are provided by densities and specific heats measured from 0.1 MPa to high pressures. During construction of the representation, appropriate regularization (minimization of specified derivatives to make thermodynamic surfaces "smooth") and use of diamond anvil cell (DAC) sound speeds allowed extension of the representational domain beyond coverage of the ultrasonic measurements, up to a pressure of 2300 MPa (near the liquid – ice VI – ice VII triple point) for temperatures between 240 to 500 K. Extending the representation to the lowest possible temperature is motivated by the interest in fundamental properties of water in the supercooled regime (*e.g.* see discussion in Holten et al., 2014) as well as a need to represent water in aqueous solutions that exhibit a significantly lowered freezing point.

Current sound speed measurements are combined with those of Lin and Trusler (2012) and higher-pressure DAC measurements from Abramson and Brown (2004), Decremps et al. (2006) and Sanchez-Valle et al. (2013). Densities and specific heats at 0.1 MPa, necessary to our determination of the properties at higher pressures, are informed by the critical reviews of data undertaken in the creation of IAPWS-95 (Wagner and Pruß, 2002) and the low temperature equation of state of Holten et al. (2014). The final equation of state was also optimized to match densities reported to 100 MPa by Kell and Whalley (1975) and to 400 MPa at low temperature by Sotani et al. (2000) and Asada et al. (2002). Further optimization of the representation at high pressure is guided by the specific heat measurements of Sirota et al. (1970), Abramson and Brown (2004), and Czarnota (1984). Other datasets used in construction of IAPWS-95 and the Holten et al. (2014) equation of state are not currently used but detailed comparisons with extant data are found in the respective publications.

In outline, the steps leading to a Gibbs energy representation are: (1) determine an LBF representation of measured sound speeds, then use a predictor-corrector algorithm based on Equations 3, 4, and 5 of Brown (2018) to directly obtain densities, specific heats, and Gibbs energy on a grid of pressures and temperatures, (2) determine by collocation a trial LBF Gibbs energy representation using linear optimization (Equation 2 in Brown (2018)), (3) modify weights and regularization until a representation is obtained that adequately matches all measurements including densities and specific heats, and (4) explore properties of the representation using standard statistical techniques.

### 3.2. The constraining measurements

#### 3.2.1. Densities and specific heats at 0.1 MPa

Holten et al. (2014) selected low temperature thermodynamic data and provided their equation of state as an alternative to IAPWS-95 in a regime below 320 K. Densities at 0.1 MPa down to 240 K from these two equations of state agree to a few tens of ppm: the IAPWS-95 values were adopted for this study. However, as show in Figure 5, low temperature specific heat measurements from various studies show systematic deviations of several percent. The specific heats calculated from IAPWS-95 and from Holten at al. (2014) follow more closely the measurements of Angell et al. (1982) and Archer and Carter (2000). This leads to systematic misfits of other measurements, including the work to higher pressures of Sirota (1970). A new representation for temperature dependence of $C_p$ below 300 K, influenced by the preponderance of other reported measurements (Sirota et al., 1970; Anisimov et al., 1972; Bertolini et al., 1985; Tombari et al., 1999) is used in the current work (Figure 5). This fit matches specific heats predicted by IAPWS-95 for temperatures greater than 300 K, while deviating from both IAPWS-95 and Holten et al. (2014) at lower temperatures to better follow the trend of measurements.

#### 3.2.2. Sound speed selection

Ultrasonic sound speeds (901 measurements in the current dataset, 210 measurements from Lin and Trusler (2012)) and DAC sound speeds behind the current equation of state are shown in Figure 6 along with a surface based on predictions from the final Gibbs energy representation. The DAC measurements (Abramson and Brown, 2004; Decremps et al., 2006; Sanchez-Valle et al., 2013), although providing constraints in the higher temperature and pressure regime, are less accurate than the ultrasonic work. Their intrinsic sound speed uncertainty is about 0.2%. However, the total uncertainty is dominated by uncertainties in pressure (about 20 MPa near room temperature and greater than 40 MPa at the highest temperatures). This will be further discussed below with the results.

The series of melting curves associated with the ice phases (Ih, III, V, VI and VII) complicate the establishment of the equation of state at the lower temperatures, particularly at low pressure. Sound speed measurements have been obtained at 0.1 MPa in the supercooled regime, down to 244 K, with an estimated uncertainty of 0.7% (Holten et al., 2014). Additional measurements are confined to the stable liquid regime at higher pressures. This leaves a triangular region below the melting curve of ice Ih, a "metastable wedge", devoid of the sound speed data needed to construct a sound speed surface. We rely on the equation of state of Holten et al. (2014), in which this gap is addressed, to fill this "metastable wedge" with provisional sound speed values. Near the melting curve of ice Ih at 250 K and 200 MPa, the sound speed values from the Holten et al. equation of state differ from our measurements by a maximum of about 500 ppm. This discrepancy is larger than our expected accuracy, but is well within that of the Holten et al. equation of state (1000 ppm above the melting curve, 5000 ppm below it). To avoid a discontinuity in the initial sound speed surface, the values from the Holten et al. equation of state were adjusted to match our data along the ice Ih melting curve. The required correction along the melt curve is linear in temperature, from 0 ppm at 273 K to 500 ppm at 250 K and 200 MPa. These modified values were only used to produce the initial sound speed surface affecting the initial Gibbs representation; the speed of sound of our equation of state was then adjusted to better match the density measurements available in the supercooled regime (including in the "metastable wedge") during optimization.

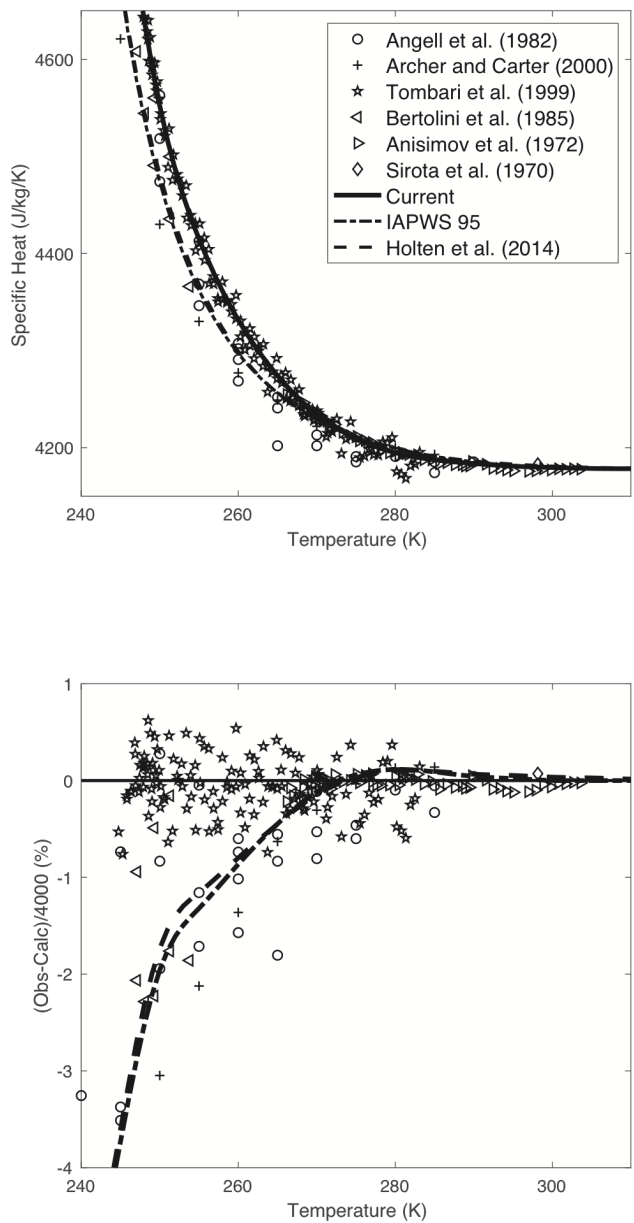

*Figure 5. Specific heat of water at 0.1 MPa. Top panel: specific heat as a function of temperature from 240 K to 310 K. Bottom panel: deviations of specific heat from the current LBF fit. Data sources are described in the legend.*

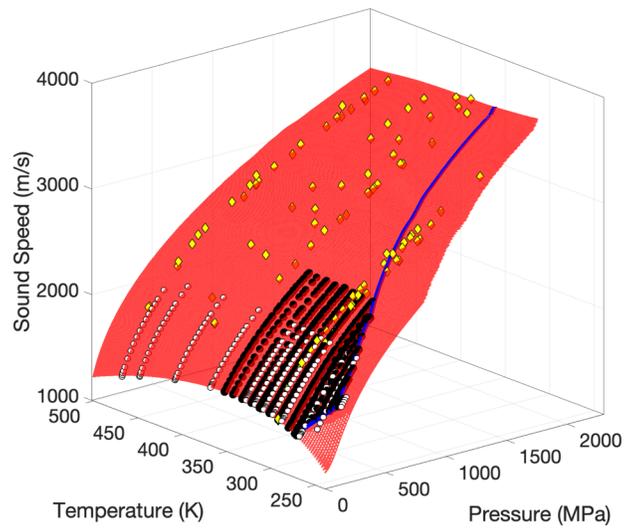

*Figure 6. Sound speed of water and predictions based on the LBF Gibbs energy representation as a function of pressure and temperature. Black circles: current ultrasonic measurements. White circles: ultrasonic measurements of Lin and Trusler (2012). Yellow diamonds: DAC sound speed measurements (see text for references). Sound speeds modified from the equation of state of Holten et al. (2014) are shown in the low-temperature, low-pressure metastable corner using small white circles. The blue lines denote the melting curves of water ices Ih, III, V, VI and VII. The surface extends 50 K into the metastable regime.*

### 3.3. Model construction

An ensemble of Gibbs energy representations was created based on differing criteria. These include efforts (1) to replicate measured sound speeds within their estimated uncertainties while also matching selected densities and specific heat measurements, (2) to explore uncertainties in derived parameters based on random errors in sound speeds, and (3) to investigate the impact of possible systematic pressure and temperature errors on determinations of density and specific heat. An optimal Gibbs energy surface has adequate complexity to successfully predict measurements, while containing variations of smoothness (i.e. "features") only as required by the underlying data. The degree of smoothing (regularization) is enforced by the size of "damping parameters" relative to the weighting of the data.

The LBF parameters of our final Gibbs energy representation (equation of state) are provided with MATLAB routines for the calculation of the thermodynamic properties (Supplementary Material C). These properties can also be evaluated using functions/subroutines available in standard numerical libraries of other programming languages. Fortran and Python packages from the authors are available on demand.

## 3.4. Results

In Figure 7, the sound speed predictions of our final equation of state (*i.e.* a Gibbs energy LBF representation) are compared to two ultrasonic datasets and two representations: the current sound speeds, those reported by Lin and Trusler (2012), and the predicted sound speeds of the IAPWS-95 and Holten et al. (2014) equations of state. The deviations are shown as a function of pressure for the isotherms along which sound speeds were measured. A body of other published sound speed measurements evaluated by Wagner and Pruß (2002) and by Holten et al. (2014) are not shown here. As documented in those two references, large (>> 1000 ppm) systematic errors for other sound speeds studies are common.

The *rms* misfit for all ultrasonic sound speeds from the current Gibbs energy representation is 100 ppm, consistent with our $2\sigma$ uncertainty estimate of 200~300 ppm (Section 2.2.4). The IAPWS-95 sound speeds at 0.1 MPa between 290 K and 370 K, where the most accurate sound speeds are available (Fujii and Masui, 1993), are matched within their stated accuracy of 50 ppm. Between 305 K and 317 K up to 300 MPa, the current representation matches the high-pressure IAPWS-95 sound speeds within about 100 ppm. The high-pressure predictions of IAPWS-95 at other temperatures show deviations from the current measurements often an order of magnitude larger. In regions of data overlap, the Lin and Trusler sound speed deviations (using the pressures based on their post-acquisition calibration) either agree with the current work or parallel the current work with offsets within 200 ppm. The Holten et al. (2014) equation of state, constructed to represent the Lin and Trusler dataset, closely follows measurements based on the published pressures.

In Figure 8, sound speed deviations from the current representation as a function of pressure are shown for three temperature ranges encompassing the DAC measurements. All deviations based on ultrasonic sound speeds, with an *rms* misfit of 0.01%, lie within the thickness of the solid lines extending to 700 MPa. The scatter in DAC measurements (within one publication and between the three reported datasets) is close to 1%. A plausible high pressure uncertainty for the sound speed surface is about 0.5%. This uncertainty will be used in Section 4.3.2 to evaluate the uncertainty of derived thermodynamic properties at high pressure.

Over most of the pressure-temperature domain, high-pressure thermodynamic properties were tightly constrained by the sound speed measurements and did not change significantly with exploration of regularization and data weights. However, differing choices for the behavior of sound speeds in the low temperature metastable wedge created differences in the low temperature equation of state and how well the representations matched the (lower-pressure) densities of Sotani et al. (2000) and the (higher-pressure) densities of Asada et al. (2002). Holten et al. (2014) assigned uncertainties of up to 300 ppm for the lower-pressure work and 1000 ppm for the higher-pressure measurements. If IAPWS-95 sound speeds are assumed in the metastable wedge, the resulting equation of state significantly misfits the Asada et al. densities. In contrast, the perturbed version of the Holten et al. sound speeds used in the current work shows a similar pattern of misfit to that reported by Holten et al. (as further documented in Section 4).

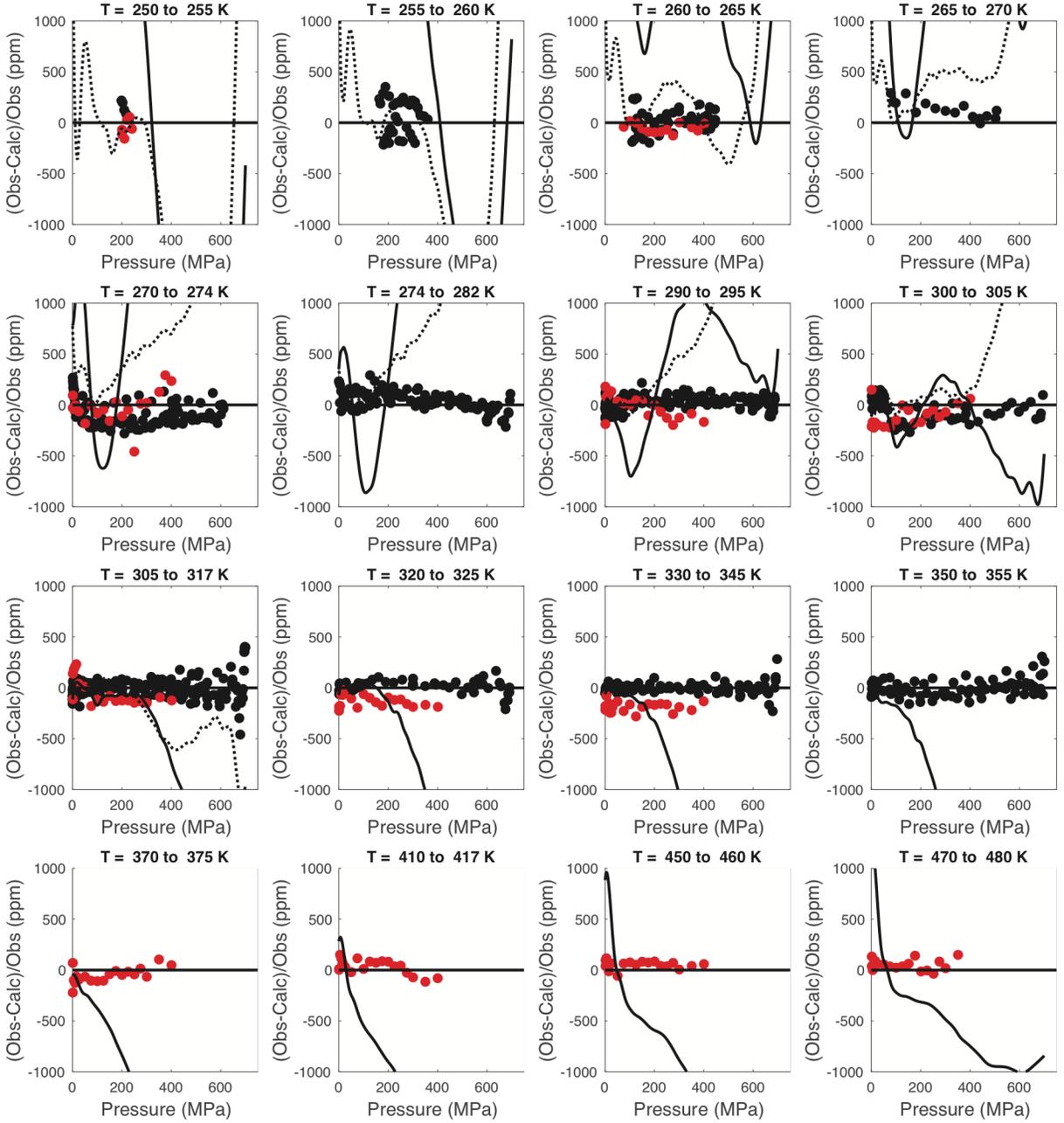

*Figure 7. Deviations of sound speeds from predictions of the LBF Gibbs energy representation. Black circles: this study. Red circles: Lin and Trusler (2012) (modified pressures). Solid line: predictions of IAPWS-95. Dotted line: predictions of the Holten et al. (2014) equation of state.*

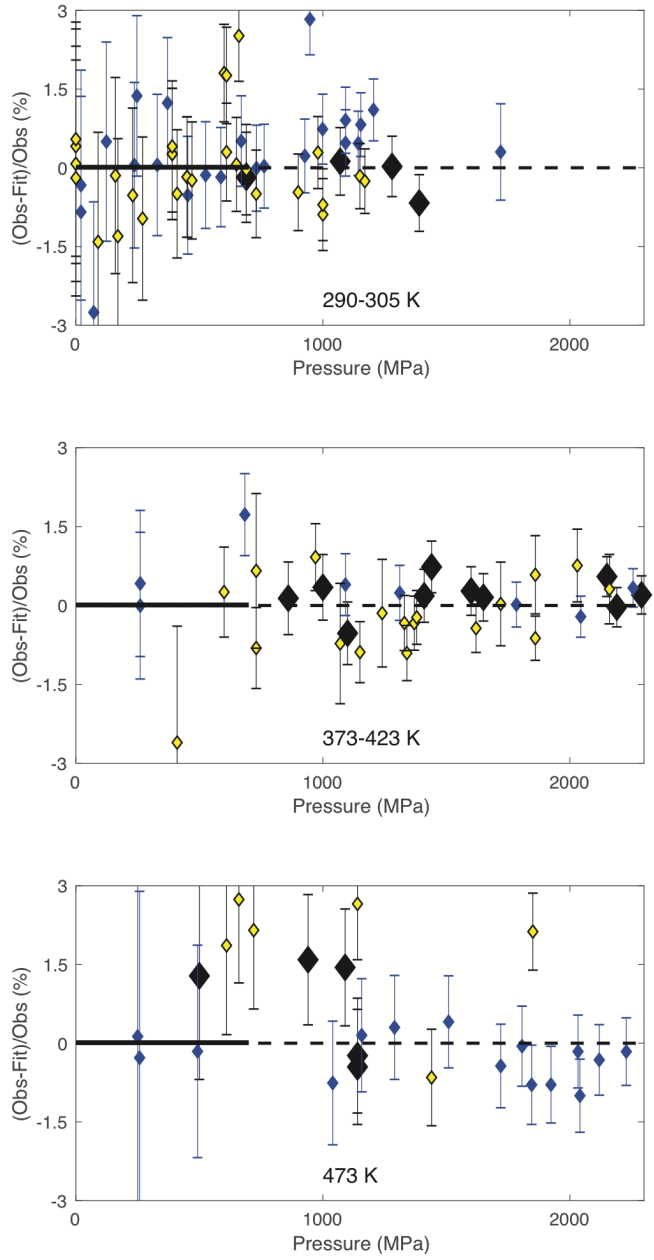

*Figure 8. Deviations of DAC sound speeds from the LBF Gibbs energy fit. Ultrasonic measurements have scatter comparable to the width of the solid lines shown below 700 MPa. Black diamonds: Abramson and Brown (2004). Blue diamonds: Decremps et al. (2006). Yellow diamonds: Sanchez-Valle et al. (2013). The plotted error bars are based on propagation of independent uncertainties for sound speed and pressure measurements.*

In Figure 9, selected thermodynamic surfaces based on the LBF representation are shown. The pressure-temperature locations of the ice melting curves are indicated on each surface. The surfaces exhibit a regular and smooth behavior both in the stable and upper metastable regimes. The specific heat surface presents the largest structures. At 0.1 MPa, as a function of temperature, the specific heat has a shallow minimum near 309 K, increasing both at higher temperatures towards the critical point at 647 K and 22 MPa, and at lower temperatures towards the proposed second critical point near 227 K (see discussion in Holten et al., 2014). At higher pressures, above 300 K, this complex behavior transitions to a featureless surface decreasing modestly with both pressure and temperature. At the lowest temperatures at all elevated pressures, the specific heat surface rolls off, trending towards its thermodynamic limit at absolute zero. Although no measurements constrain the equation of state in the deeply metastable regimes at low temperature, the behavior of the surfaces in these regions of extrapolation are reasonable.

# 4. Discussion

## 4.1. Density

The accuracy of density measurements for water at 0.1 MPa has been evaluated by Tanaka (2001). His values for air-saturated water with an uncertainty of 1 ppm are enforced in the current Gibbs representation. Literature reports of densities at higher pressure were evaluated by Wagner and Pruß (2002) and by Holten et al. (2014). Although a large number of measurements have been reported, a subset of sources provides the largest influence on the construction of the equations of state for water. Here, the final Gibbs energy representation was optimized against three primary high-pressure datasets. Kell and Whalley (1975) reported densities to 100 MPa from 300 to 450 K, with uncertainties estimated in the range of a few tens of ppm. Sotani et al. (2000) and Asada et al. (2002) reported densities from 253 to 298 K at pressures to 400 MPa; both low temperature studies were accomplished in the same laboratory. A digitized version of their data is provided in the supplemental material of Holten et al. (2014). The relationship of the three high-pressure datasets to a broader selection of reported densities can be viewed in both Wagner and Pruß (2002) and in Holten et al. (2014).

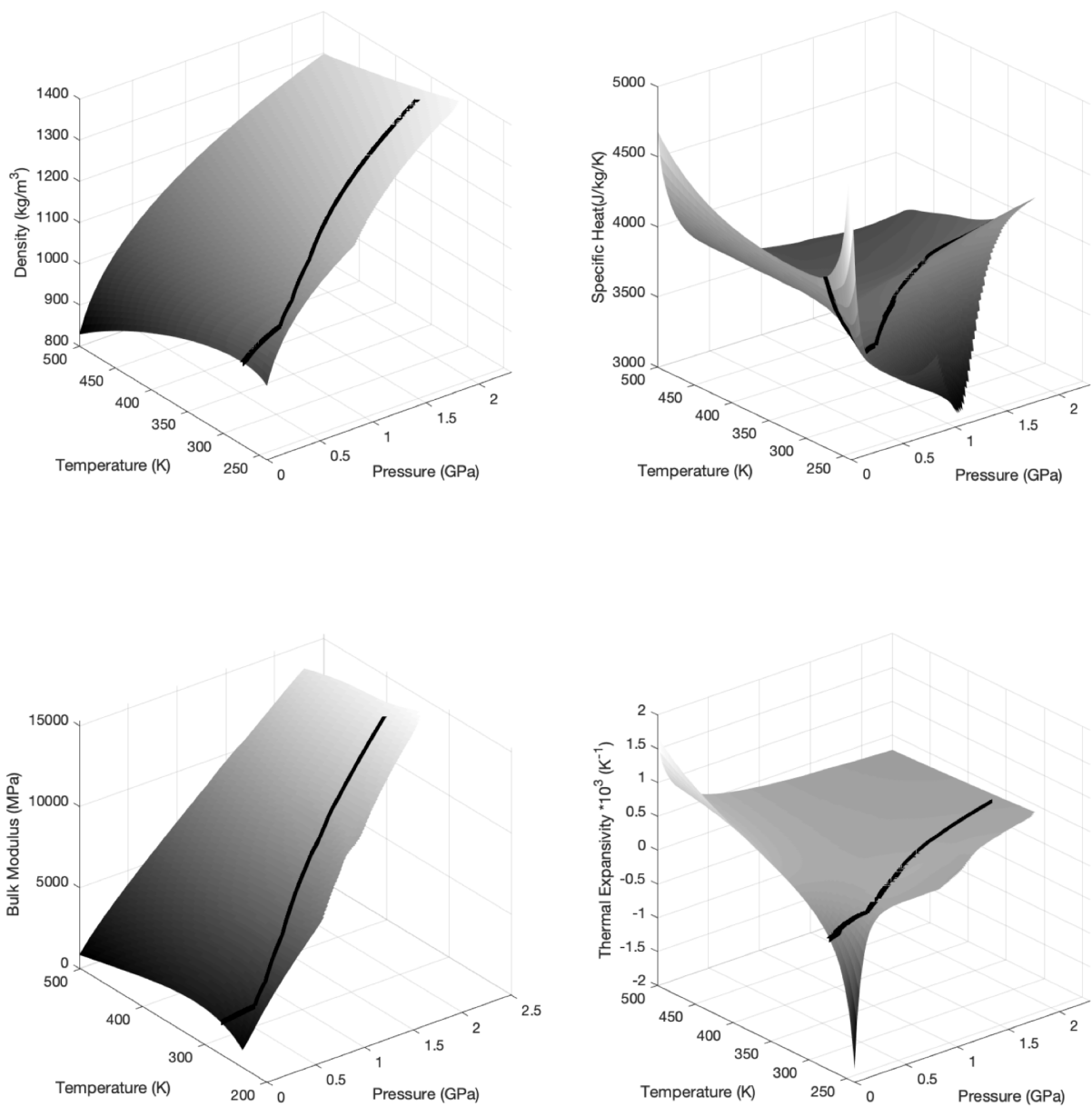

*Figure 9. Thermodynamic properties of water from 240 K to 500 K, between 0.1 and 2300 MPa, based on the current LBF Gibbs energy representation. Top left panel: density. Top right panel: constant pressure specific heat. Bottom left panel: isothermal bulk modulus. Bottom right panel: thermal expansivity. On each surface, the black lines represent the projections of the pressure-temperature coordinates of the melting curves of the ice polymorphs (ices Ih, III, V and VI). At higher pressures, metastable properties based on extrapolation are shown to temperatures 50 K below melting.*

In Figure 10, deviations of these selected densities from the current Gibbs energy representation are shown. At 300 K, densities predicted by both the IAPWS-95 and Holten et al. (2014) equations of state agree with densities derived from the current representation and with the Sotani et al. (2000) and Asada et al. (2002) high-pressure densities to within 200 ppm. With decreasing temperature, the IAPWS-95 densities show deviations far exceeding the reported uncertainty for the Sotani et al. and Asada et al. measurements. At 253 K and 400 MPa, where no data constrained the IAPWS-95 formulation, densities deviate by over 2000 ppm (0.2%). Densities based on Holten et al. closely agree with the current work. The Sotani et al. and Asada et al. densities show deviations within 300 ppm, close to their estimated uncertainties.

## 4.2. Specific heat

In Figure 11, isobars for the constant pressure specific heat $C_p$ are shown as functions of temperature with higher-pressure isobars offset to improve legibility. Specific heat decreases with increasing pressure; the high $C_p$ values at 0.1 MPa at the lowest temperatures are rapidly suppressed by pressure. On each isobar, the melting point (black star) is marked to separate the regime of measurements from that of extrapolation into the metastable regime. As a result of the thermodynamic contribution from a proposed liquid-liquid transition, the low-temperature equation of state of Holten et al. (2014) predicts values of $C_p$ larger than those of IAPWS-95, with larger differences at higher pressures. The current equation of state follows closely the predictions of Holten et al. in the stable domain constrained by the ultrasonic measurements, before transitioning to a fall-off delayed but similar to IAPWS-95 in the metastable domain (the equation of state is constrained in the metastable domain by reference densities up to 400 MPa). These positive deviations to IAPWS-95 shift to higher temperatures with increasing pressure, and are constrained by the measured sound speeds at pressures up to 700 MPa.

In Figure 12, the pressure dependence of $C_p$ is shown for isotherms at 300 K and 384 K. Obvious for both isotherms and measurements of Sirota et al. (1970) is the significant decrease of $C_p$ with pressure in the low-pressure regime. For the 384 K isotherm, the current representation, IAPWS-95, and measurements of Abramson and Brown (2004) agree to about a percent and show a continued decrease in $C_p$ at the highest pressures. At 300 K, the IAPWS-95 isotherm shows a monotonic decrease in $C_p$ with pressure. In contrast, the current representation shows a minimum near 400 MPa followed by a marked increase to the melt line (about 1100 MPa at 300 K). The measurements of Czarnota (1984) and Abramson and Brown (2004) support the trend of the current representation. Experimental uncertainty near 1% for $C_p$ likely is responsible for the misfit of the data by the representation.

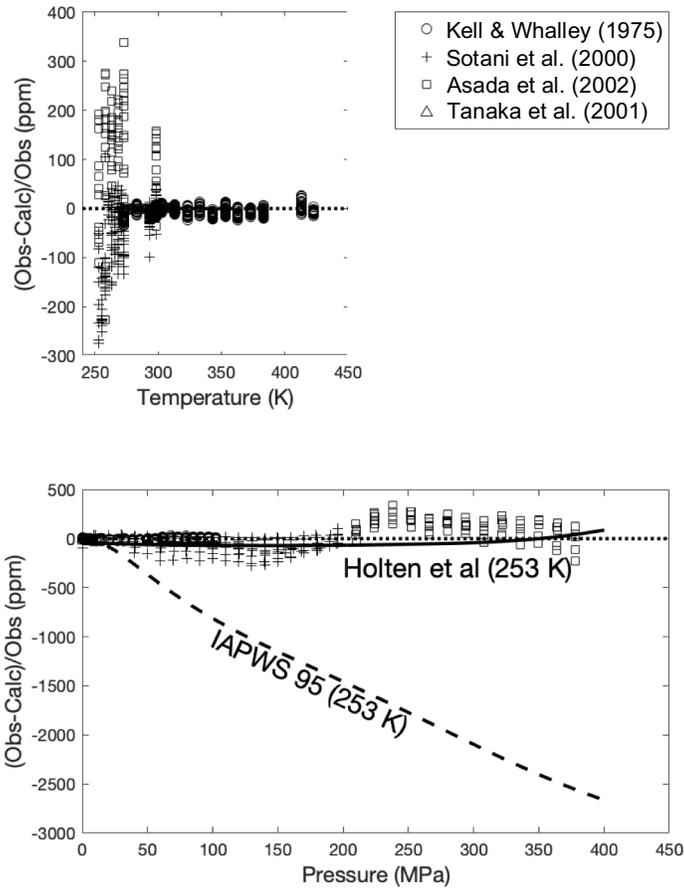

*Figure 10. Density deviations from the LBF Gibbs energy representation. Upper panel: deviations as a function of temperature for all pressures. Lower panel: deviations as a function of pressure for all temperatures. Sources of measured densities are listed in the legend. Deviations of density predicted by IAPWS-95 and Holten et al. (2014) are shown only for the 253 K isotherm.*

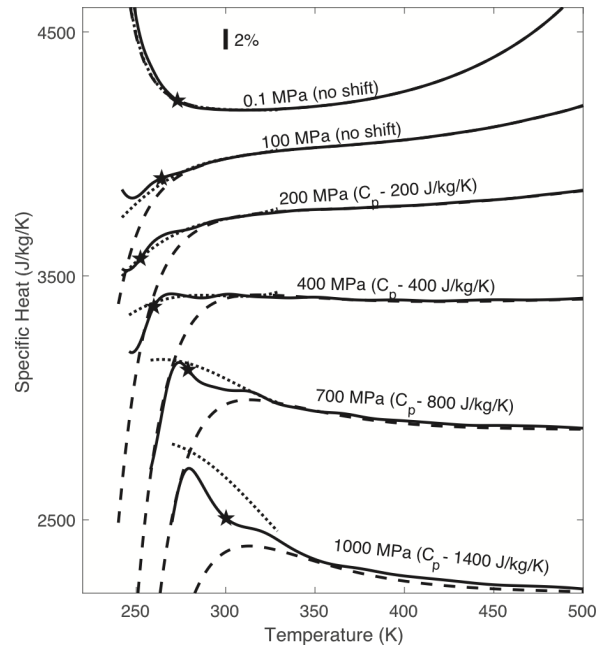

*Figure 11. Constant pressure specific heat ($C_p$) as a function of temperature for selected isobars (pressures are labeled for each isobar). A bar representing 2% differences in $C_p$ is shown. The isobars are shifted vertically (as indicated by values in parentheses) to avoid overlap of curves. Solid lines are predictions of current analysis. Dashed lines are IAPWS-95 values. Dotted lines are based on Holten et al. (2014). All curves extend into the metastable regime at low temperatures; stars indicate the melting point for each isobar.*

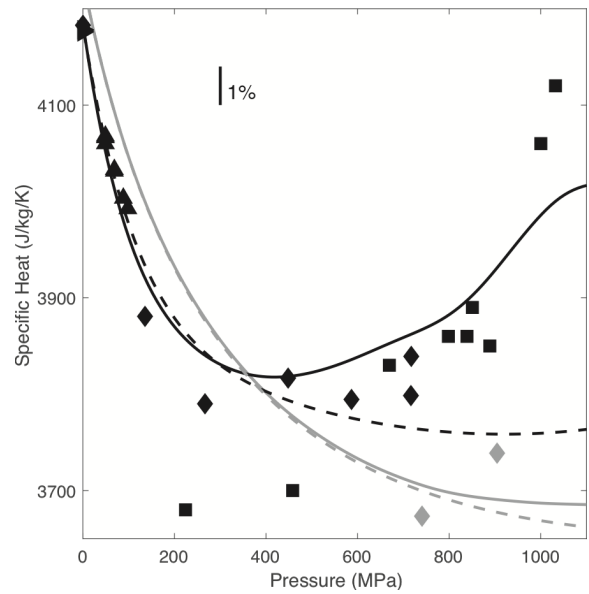

*Figure 12. Constant pressure specific heat ($C_p$) as a function of pressure at 300 K (black symbols) and 384 K (gray symbols). Solid line is predictions of current analysis. Dashed line is IAPWS-95. Measurements at ambient temperature (298-300 K) are shown as black triangles (Sirota et al., 1970), black diamonds (Abramson and Brown, 2004), and black squares (Czarnota, 1984). The gray diamonds at 384 K are from Abramson and Brown (2004). A bar representing a 1% difference in $C_p$ is shown.*

## 4.3. Error analysis

### 4.3.1. Random errors in sound speed

Uncertainties in heat capacities and densities associated with random errors in the ultrasonic sound speed measurements were estimated through a Monte Carlo analysis. For that process, the Lin and Trusler (2012) and current ultrasonic measurements are replaced with predictions from our final equation of state (at the same pressure-temperature coordinates), random errors are added (with σ = 100 ppm, equivalent to the observed scatter of the measurements), and a new equation of state is produced following the steps previously undertaken with the real ultrasonic datasets. The exercise is repeated (each time with a different random distribution of errors), and the results are compared with the reference model. This process offers an insight on how the random errors associated with the sound speeds affect the equation of state. Owing to the density of data of the ultrasonic datasets, the equation of state remains mostly unaffected by this process. Density deviations remain essentially at the ppm level, with a local increase to 20 ppm near the melting curves. Higher deviations up to 100 ppm are observed in the metastable domain. Throughout almost all the stable region above the melting curves, specific heat deviations remain at or below 0.1%, a value equivalent to the accuracy of the reference heat capacity measurement at 0.1 MPa. The only exception is observed around the melting curve of ice VI, with deviations reaching 1%.

### 4.3.2. Systematic errors

The adopted temperature uncertainty of 50 mK appears to have a small to negligible impact on derived thermodynamic properties. We systematically (both linearly and quadratically with temperature) and randomly altered measured temperatures within uncertainties and found that propagated uncertainties in specific heat remained below 0.1% (the uncertainty at 0.1 MPa). Thus, temperature uncertainty is currently ignored in our analysis.

To propagate systematic errors in pressure, the pressures associated with the sound speeds were altered per our estimate of pressure or sound speed uncertainty. A relative pressure uncertainty of 0.04% was adopted within the range of ultrasonic measurements (to 700 MPa below 353 K, and to 400 MPa above 353 K where only the Lin and Trusler (2012) dataset is available). At pressures beyond the domain of the ultrasonic measurements, the DAC experiments have substantially lower pressure accuracy. Based on examination of Figure 8, we estimate the individual DAC sound speeds to be accurate to about 1% and a surface derived from the ensemble collection of measurements to be accurate within about 0.5%. This is equivalent to a pressure error of about 10 MPa at 700 MPa and 30 MPa at 2300 MPa. Pressure uncertainties were thus allowed to increase linearly with pressure, from 0.04% at the highest ultrasonic pressures to 0.5% of sound speed near 1000~1500 MPa and beyond.

Figure 13 provides a contoured summary of propagated errors based on the estimated pressure errors. Density errors are limited to 25 ppm in the domain of ultrasonic measurements and grow to 1400 ppm at 2300 MPa. For specific heat, below 1000 MPa, uncertainties from the systematic pressure errors do not exceed 0.1% (the accuracy of the reference values at 0.1 MPa); at 2300 MPa, uncertainties reach 0.4 to 0.7%. The sound speed deviations are consistent with the assumed systematic uncertainties used to construct the modified equation of state. In the domain of our ultrasonic measurements, the 0.04% uncertainty in pressure translates into a systematic error of 0.02% in apparent sound speed. We believe this value represents a reasonable estimate of the overall accuracy of our sound speed measurements.

### 4.3.3. Deviations of current representation from IAPWS-95

Deviations from IAPWS-95 of densities, specific heats and sound speeds are summarized in Figure 14. In line with the results shown in Figure 4, deviations of the sound speeds in the domain of ultrasonic measurements remain within 0.4%. The DAC constraints result in systematic deviations of sound speed reaching -1.2% at 2300 MPa. As a result of the changing sign of sound speed deviations (positive in the low-pressure regime and negative beyond about 1000 MPa), density deviations remain relatively small, reaching 0.1~0.3% only near the melting curves. This region is especially well constrained by both direct density measurements and a high density of accurate sound speed measurements from this study. The largest deviations of specific heat (close to 20%, in agreement with the IAPWS R6-95 estimates established after the study of Lin and Trusler (2012)) are also found in this regime of low temperatures and pressures.

## 5. Conclusion

A new equation of state of liquid water at sub-critical temperatures, at pressures up to 2300 MPa (covering the triple point with ices VI and VII) was created using new sound speed measurements combined with previously published thermodynamic data. Our new dataset, obtained from six separate loads, is comprised of 901 individual measurements acquired along isotherms between 253 and 353 K, at pressures from 0.1 to

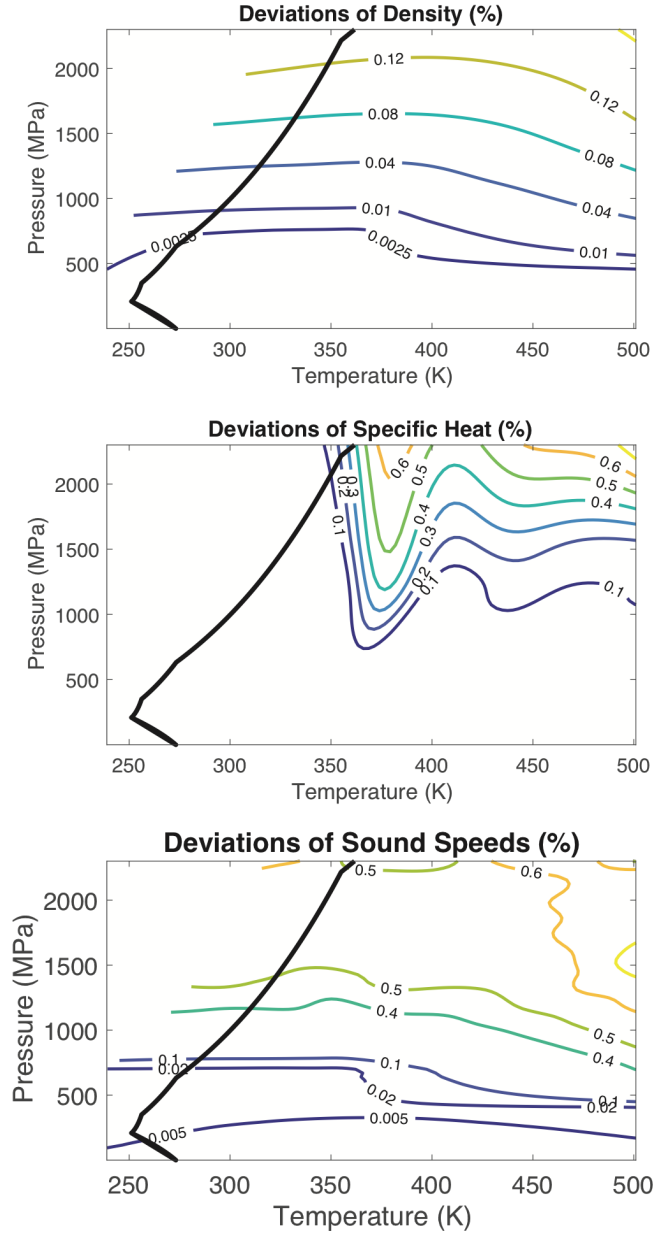

*Figure 13. Errors in derived thermodynamic properties based on systematic errors in pressure. Upper panel: density deviations. Middle panel: specific heat deviations. Lower panel: sound speed deviations. The melting curves are indicated with the thick black lines. Within the range of the ultrasonic data, the pressure error is 0.04%. Pressure uncertainties reach about 30 MPa at 2300 MPa in the domain of the DAC measurements.*

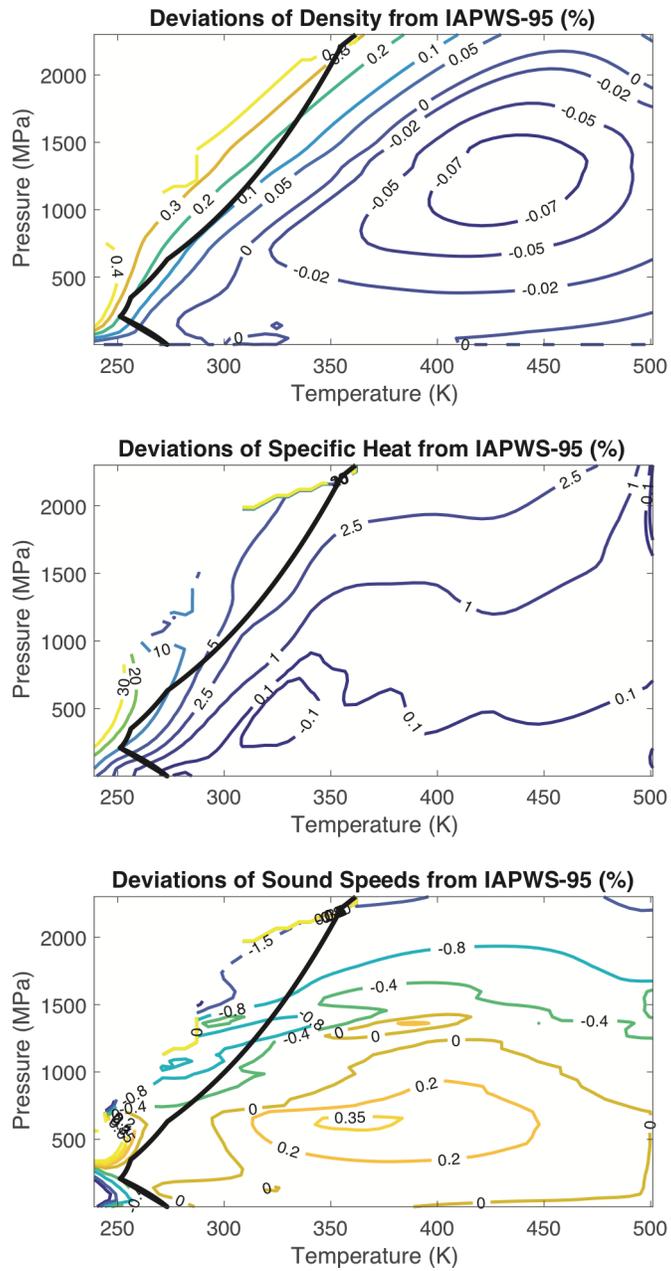

*Figure 14. Deviations of the LBF Gibbs energy representation from IAPWS-95. Upper panel: density deviations. Middle panel: specific heat deviations. Lower panel: sound speed deviations. The melting curves are indicated with the thick lines.*

700 MPa. Over the entire domain explored, the 2σ precision and accuracy of our ultrasonic sound speed measurements are close to 0.02%. This represents an improvement of an order of magnitude relative to previous works reported above 400 MPa. Below 400 MPa, our dataset closely matches the ultrasonic measurements of Lin and Trusler (2012). The agreement between the two studies, with similar precision but relying on independent determinations of absolute pressure, supports the observed systematic deviations from the IAPWS-95 standard.

The precision with which speeds of sound are measured at high pressures makes it a robust constraint for refinement of a Gibbs energy representation. Using the local basis function (LBF) framework of Brown (2018), our results were combined with other published measurements to produce a new equation of state for pure water. The new representation builds on the work of Holten et al. (2014) in the metastable region below the melting curves and of IAPWS-95 at 0.1 MPa at higher temperatures. The dataset of Lin and Trusler (2012) (to 400 MPa and 473 K) and our new measurements (to 700 MPa and 353 K) underlie the LBF equation of state at intermediate pressures. The regime at higher pressures was constrained using sound speeds determined in diamond anvil cells (Abramson and Brown, 2004; Decremps et al., 2006; Sanchez-Valle et al., 2013). The new representation covers a large part of the subcritical domain, extending from 240 to 500 K, and from 0.1 to 2300 MPa.

The global accuracy of our equation of state was assessed through the exploration of selected random and systematic errors. As a result of the high density of data available with the Lin and Trusler (2012) and current ultrasonic datasets, random errors in sound speeds have a negligible impact on the LBF representation. The impact of the systematic errors is limited as well, owing in part to the achievable accuracy of measurements. The main contribution limiting the accuracy of the new equation of state is the relatively lower accuracy of diamond anvil cell pressure measurements above 700 MPa. The largest improvements to the IAPWS-95 formulation are observed above 100 MPa, at temperatures near the melting curves of ices Ih, III and V: the accuracy of the density representation is improved from 0.1-0.2% to better than 0.01%, and the specific heat from 5-20% to better than 1%. For the specific heat, the precision in changes of specific heat with pressure (0.03%) is higher than the absolute accuracy of the extant 0.1 MPa measurements, limiting the improvements at higher pressure.

The accuracy of the new sound speed measurements supports the large positive deviation to IAPWS-95 observed in the new specific heat representation. The deviation increases in intensity and shifts to slightly higher temperatures at high pressure. In the domain covered by ultrasonic measurements, the deviation coincides in pressure and temperature with the effect on specific heat of a suggested liquid-liquid transition incorporated in the Holten et al. (2014) equation of state.

The local basis function framework adopted for the current equation of state will make possible incremental improvements, based either on measurements or theory, and will allow its extension in pressure and temperature without impact on the representation within the current domain. With pure water as a compositional end-member, the present work offers the prospect of flexible equations of state for aqueous solutions matching or exceeding IAPWS-95 levels of accuracy.

## Supplementary Material

A more detailed account of the pressure, temperature and sound speed measurements behind the ultrasonic dataset acquired for this study is provided in Supplementary Material A. The acquired sound speed dataset in presented as an ASCII file in Supplementary Material B. The LBF parameters of the equation of state, the thermodynamic datasets used as references in this study, and MATLAB scripts to calculate the thermodynamic properties from the LBF parameters are combined in Supplementary Material C.

## Acknowledgements

Funding for this work was provided by NASA Solar System Workings Grant 80NSSC17K0775 and by the Icy Worlds node of NASA's Astrobiology Institute (08-NAI5-0021). Effort of the following students as part of an undergraduate research experience is appreciated: Guy Giesa-Wilson, Penny Espinoza, Nathan Reinsdorf, Jason Ott, Sarah Newport, and Fred Bradley. In addition, Mr. Giesa-Wilson and Ms. Espinoza wrote valuable code used in ongoing data acquisition and analysis. Continuing discussions with S. Vance, E. Abramson, and B. Journaux has improved the outcome of this work. We appreciate the willingness of M. Trusler to share details of the pressure calibration used in the earlier study. We are thankful for the assistance of William Wall of Harwood Engineering in obtaining an accurate pressure calibration. We thank Frederic Decremps for providing the data tables associated with their study. O. Bollengier is grateful to